\documentclass[dvips]{article}

\usepackage{icrc2011}

\title{Discovery of VHE gamma-ray emission from the direction of the globular cluster Terzan~5}

\newcommand{\etal}{\MakeLowercase{\textit{et al. }}} 
\shorttitle{Domainko \etal VHE gamma-rays from Terzan~5}

\authors{Domainko, W.$^{1}$, Clapson, A.-C.$^{1,2}$, Brun, F.$^{3}$,
Eger, P.$^{4}$, Jamrozy, M.$^{5}$, Dyrda, M.$^{6}$, Komin, N.$^{7}$, Schwanke, U.$^{8}$ for the H.E.S.S. collaboration}
\afiliations{$^1$Max-Planck-Institut fuer Kernphysik, P.O. Box 103980, D 69029
Heidelberg, Germany
\\$^2$EIPOD fellow, Cell Biology and Biophysics \& Developmental Biology Units, EMBL Heidelberg, Meyerhofstrasse 1, 69117 Heidelberg
\\$^3$Laboratoire Leprince-Ringuet, Ecole Polytechnique, CNRS/IN2P3, F-91128 
Palaiseau, France
\\ $^4$Universit\"at Erlangen-N\"urnberg, Physikalisches Institut, Erwin-Rommel-Str. 1,
D 91058 Erlangen, Germany
\\ $^5$Obserwatorium Astronomiczne, Uniwersytet Jagiello{\'n}ski, ul. Orla 171,
30-244 Krak{\'o}w, Poland
\\ $^6$Instytut Fizyki J\c{a}drowej PAN, ul. Radzikowskiego 152, 31-342 Krak{\'o}w,
Poland 
\\ $^{7}$Laboratoire d'Annecy-le-Vieux de Physique des Particules,
Universit\'{e} de Savoie, CNRS/IN2P3, F-74941 Annecy-le-Vieux,
France
\\ $^8$Institut f\"ur Physik, Humboldt-Universit\"at zu Berlin, Newtonstr. 15,
D 12489 Berlin, Germany}
\email{wilfried.domainko@mpi-hd.mpg.de}

\abstract{Globular clusters are old stellar systems which exhibit very-high stellar densities in their cores. The globular cluster Terzan 5 is characterized by a high stellar encounter rate and hosts the largest detected population of millisecond pulsars. It also features bright GeV gamma-ray emission and extended X-ray radiation. However, no globular clusters have been detected in very-high-energy gamma rays (VHE, E$>$ 100 GeV) so far. In order to investigate this possibility Terzan 5 has been observed with the H.E.S.S. telescope array in this energy band. The discovery of a source of VHE gamma rays from the direction of this globular cluster will be reported. The results of the VHE analysis and a multi-wavelength view of Terzan 5 will be presented in this contribution. No counterpart or model can fully explain the observed morphology of the detected VHE gamma-ray source.}

\begin{document}
\maketitle

\section{Introduction}

Several types of galactic VHE gamma-ray sources like pulsar wind nebulae (PWNe) and supernova remnants (SNRs) have been detected to date but none so far in the vicinity of a globular cluster. Globular clusters are old stellar systems which exhibit very-high stellar densities in their cores. This leads to numerous stellar encounters \cite{pooley2006} and to the prolific formation of millisecond pulsars (msPSRs) \cite{ransom2008}. Globular clusters are predicted VHE gamma-ray sources. In these models the gamma-ray emission is produced by inverse Compton (IC) up-scattering of stellar radiation fields and the cosmic microwave background by relativistic electrons originating either from the msPSRs themselves or their PWNe \cite{bednarek2007,venter2009,cheng2010}.

Here the discovery of VHE gamma-ray emission from the direction of the Galactic globular cluster Terzan~5 is reported. This globular cluster is located at a distance of 5.9~kpc \cite{ferraro2009}
at RA(J2000)~17$^\mathrm{h}$48$^\mathrm{m}$04$^\mathrm{s}$.85 and 
Dec~$-24^{\circ}$46$^\prime$44$^{\prime\prime}$.6 
(Galactic coordinates: $l = 3.84^{\circ}$, $b = 1.69^{\circ}$)
and exhibits a core radius $r_\mathrm{c} = 0^\prime.15$, a half-mass radius 
$r_\mathrm{h} = 0^\prime.52$ and a tidal radius $r_\mathrm{t} = 4^\prime.6$ \cite{lanzoni2010}. Terzan~5 hosts the largest population of detected msPSRs (33) \cite{ransom2008} and it has been discovered by \emph{Fermi-LAT} in the GeV range \cite{kong2010,abdo2010}. 

So far only upper limits have been reported from globular clusters in the VHE gamma-ray range (e.g. \cite{aharonian2009,anderhub2009,mccutcheon2009}).

\section{Observation and Analysis}

\begin{figure*}[ht]
\centering
\includegraphics[width=13cm]{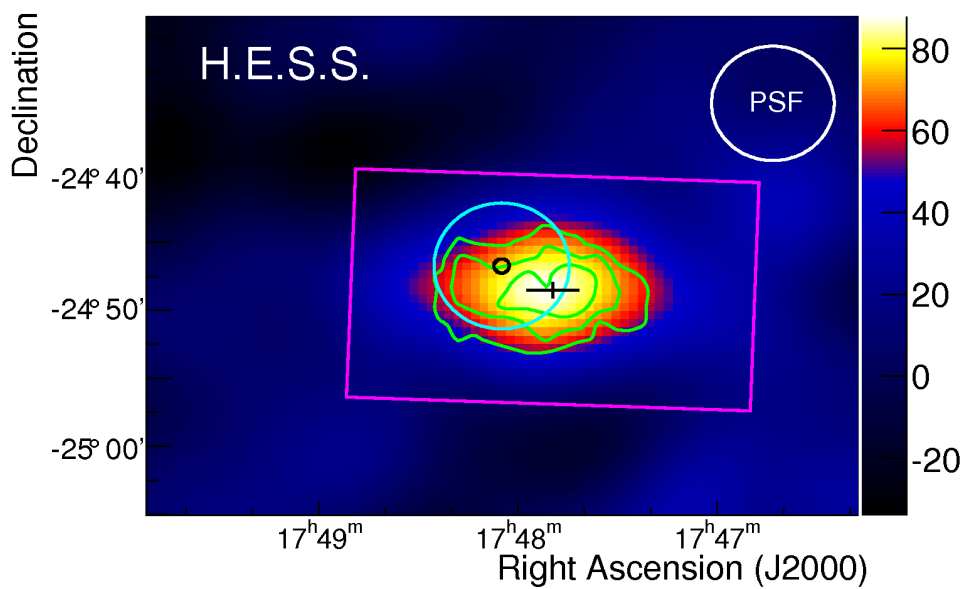}
\caption{Exposure-corrected excess image from the H.E.S.S. data,
smoothed with a Gaussian function of width 0.1$^\circ$ and
overlaid with significance contours (4 -- 6~$\sigma$) in RADec~J2000 coordinates. 
The circles show the half-mass radius (in black)
and the larger tidal radius (in cyan) of the globular cluster.
The cross indicates the best-fit source position,
assuming a 2D Gaussian shape,
with 1~$\sigma$ uncertainty on each axis.
The rectangle represents the integration
region used for the full-source spectral analysis. The upper-right corner circle illustrates the instrumental PSF.
}
\label{figure:excess_hess}
\end{figure*}

The observations presented here have been undertaken with the H.E.S.S. array which is an array of four Imaging Atmospheric Cherenkov Telescopes located in the Khomas highlands in Namibia. Stereoscopic trigger and analysis method allow efficient cosmic ray background rejection and accurate energy and arrival direction reconstruction for gamma-rays in the energy range 100 GeV - 100 TeV. Terzan 5 has been observed for 90 hours of good life-time with 3- and 4 telescopes with an average zenith angle of 20.4$^\circ$ and a mean pointing direction off-set of 0.95$^\circ$. These observations resulted in the detection of a source of VHE gamma-rays in the vicinity of Terzan~5. Using \textit{hard cuts} \cite{aharonian2006} a significance of 5.3 $\sigma$ at the position of Terzan~5 is found with a nearby peak significance of 7.5 $\sigma$. This is confirmed by an independent calibration and analysis chain \cite{denaurois2009}. The source appears to extend beyond the tidal radius of the globular cluster. A 2-dimensional Gaussian fit
results in a best-fit position of 
RA(J2000)~17$^\mathrm{h}$47$^\mathrm{m}$49$^\mathrm{s} \pm 1^\mathrm{m}.8_\mathrm{stat} \pm 1^\mathrm{s}.3_\mathrm{sys}$ and 
Dec~$-24^{\circ}$48$^\prime$30$^{\prime\prime} \pm 36^{\prime\prime}_\mathrm{stat}  \pm 20^{\prime\prime}_\mathrm{sys}$,
offset by 4$^\prime$.0 from the GC center. Therefore the source is named HESS~J$1747-248$. 
The size of the source is given by
the Gaussian widths 9$^\prime$.6$ \pm $2$^\prime$.4
and 1$^\prime$.8$ \pm $1$^\prime$.2, 
for the major and minor axes respectively, 
oriented 92$^\circ \pm $6$^\circ$ westwards from North.

For spectral analysis a more restrictive data selection has been applied to improve the energy reconstruction which resulted in 62 hours of data. For a power-law spectral model of the form $k \left( \frac{E}{E_0} \right)^{-\Gamma}$, the flux normalization $k$ at $E_0 = 1$~TeV is 
(5.2$\pm$1.1)$\times$10$^{-13}$~cm$^{-2}\,$s$^{-1}\,$TeV$^{-1}$
and the spectral index $\Gamma = 2.5 \pm 0.3_\mathrm{stat} \pm 0.2_\mathrm{sys}$. This corresponds to an integral flux of (1.2$\pm$0.3)$\times$10$^{-12}$~cm$^{-2}\,$s$^{-1}$, or 1.5\% of the Crab flux,
in the 440~GeV --- 24~TeV range. There are not enough excess events to discuss a more complex spectral model.

\begin{figure*}[ht]
\centering
\includegraphics[width=9.0cm]{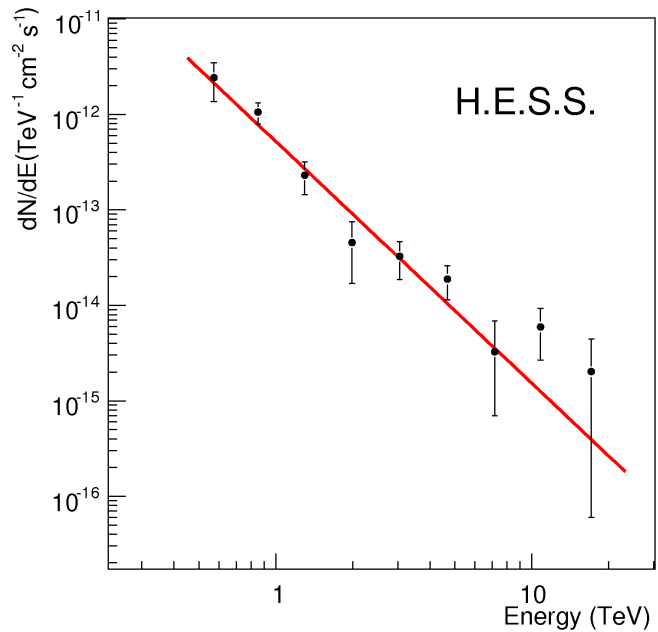}
\caption{VHE $\gamma$-ray spectrum of HESS~J1747-248
with 1~$\sigma$ error bars, fitted with a power-law model.
The fit results are discussed in the text.
}
\label{figure:spectrum_hess}
\end{figure*}

\section{Multi-wavelength environment}

Several interesting structures have been found in the surroundings of HESS~J$1747-248$ in archival multi-wavelength data. In the X-ray regime
diffuse emission extending beyond $r_\mathrm{h}$ has been reported \cite{eger2010}. This diffuse emission is centered on the core of Terzan~5, exhibits an unabsorbed flux of $(5.5 \pm 0.8) \times 10^{-13}$ erg cm$^{-2}$ s$^{-1}$ in the 1~-~7~keV band and is most likely of non-thermal origin with a hard spectrum with photon index $0.9\pm0.5$. Also diffuse radio emission has been found which extends to the north-west of Terzan~5 but does not show a tale-telling morphology like e.g. a SNR shell \cite{clapson2011}. The origin of the diffuse X-ray emission as well as the diffuse radio emission is ambiguous but could be connected to the large population of msPSRs in Terzan~5 \cite{eger2010,clapson2011}. Several scenarios for VHE gamma-ray emission would predict multi-wavelength counterparts, however, the relation between HESS~J$1747-248$ and the diffuse X-ray and radio sources remains unclear at the moment.

\section{Discussion}

Available multi-wavelength data on the one hand do not show any typical VHE gamma-ray emitter (like PWN or SNR) in the vicinity of the detected source. On the other hand the properties of the source, namely the extension and the off-set ($2 \sigma$ level) are unexpected for a globular cluster. Therefore the results of the H.E.S.S. observations are difficult to interpret. Here scenarios excluding and including the globular cluster as the origin of the VHE gamma-ray emission are briefly discussed.

\subsection{Chance coincidence}

The positional concurrence between the globular cluster and the VHE gamma-ray source could in principle be just a chance coincidence of physically unrelated objects. Notably the source parameters (extension, photon spectrum) of HESS~J$1747-248$ are compatible with the properties of VHE gamma ray detected PWNe \cite{mattana2009}. 

The probability of chance coincidence can be estimated from the distribution of VHE gamma-ray sources in the galactic disk. From the H.E.S.S. galactic plane scan the lateral distribution of sources in the longitude range from $-85^\circ - +60^\circ$ has been obtained. It can be described by a Gaussian profile (containing 48 sources) centered at b=$-0.26^\circ$ with a width of $0.4^\circ$ and additional four outliers with b $<-2^\circ$ below the galactic disk \cite{chaves2009}. At a latitude of $1.7^\circ$ Terzan~5 is almost 5 $\sigma$ away from the center of the Gaussian distribution and there are no other VHE sources detected in the latitude band of $1.5^\circ - 2^\circ$. Hence HESS~J$1747-248$ presents an outlier to the lateral source distribution. If it is assumed that one outlier is located in the latitude band of $1.5^\circ - 2^\circ$ then the chance coincidence that it is placed within $0.1^\circ$ of the center of the globular cluster is about 10$^{-4}$. Thus, its quite unlikely that the proximity of Terzan~5 and HESS~J$1747-248$ is due to chance but this possibility cannot be excluded.

\subsection{Leptonic scenario}

Globular clusters are predicted VHE gamma-ray emitters \cite{bednarek2007,venter2009,cheng2010}. All these models rely on IC up-scattering of low energy photons (stellar radiation, cosmic microwave background) by energetic electrons. The electrons are proposed to be accelerated either by the population of msPSRs themselves or by their colliding PWNe. Models \cite{bednarek2007,venter2009} predict a VHE gamma-ray flux for Terzan~5 of about 1\% of the flux of the Crab nebula for reasonable input parameters. This is in broad agreement with the flux of HESS~J$1747-248$.

For an IC scenario it is expected that the gamma-ray emission should follow the shape of the up-scattered radiation field which is highly centrally peaked in the case of a globular cluster. Therefore, an extended source with off-set from the globular cluster center is challenging to interpret for such a scenario. IC emission in the VHE gamma-ray regime should be accompanied by synchrotron emission in the X-ray regime. However, since the detected diffuse X-ray emission \cite{eger2010} is also centered on the globular cluster core a simple model where the X-ray and VHE gamma-rays emission originates from the same population of electrons cannot explain the morphology of the VHE source.

To summarize, the population of msPSRs would energetically be capable of producing the VHE gamma-ray emission but the source morphology is not self-evident for such a scenario.

\subsection{Hadronic scenario}

Globular clusters are believed to boost the rate of stellar mergers due to the extreme stellar densities in their cores (e.g. \cite{shara2002,grindlay2006}). These collisions lead in some cases to stellar explosions like SN Ia powered by white~dwarf~-~white~dwarf mergers. Remnants of these SN Ia could be the acceleration site of hadronic cosmic rays (see \cite{acero2010} for the detection of SN 1006 in VHE gamma-rays).

To explain the VHE emission an energy in hadronic cosmic rays of 10$^{51} (n/0.1\mathrm{cm}^{-3})$~erg ($n$ density of ambient target material) would be needed if it is assumed that the cosmic ray spectrum follows a power-law with index -2 below the region which can be probed with H.E.S.S. This is somewhat high for a supernova. The source morphology could be explained by a hadronic scenario but there is no multi-wavelength support for a SNR origin \cite{clapson2011} of the VHE gamma-ray emission.

\section{Summary}

A source of VHE gamma-ray emission,  HESS~J$1747-248$, has been detected in the vicinity of the Galactic globular cluster Terzan~5. It exhibits a flux of about 1.5\% of the flux of the Crab nebula above 440~GeV and a spectrum which can be described by a power-law with index $-2.5$. The source appears to be extended, off-set from the core of Terzan~5 but overlaps significantly with the globular cluster. The probability of a chance coincidence between HESS~J$1747-248$ and Terzan~5 is low but this possibility cannot be ruled out completely. If the VHE gamma-ray source is indeed physically connected with the globular cluster it would represent the first member of a new type of VHE gamma-ray emitters. The nature of the source is still uncertain. On the one hand its properties (e.g. morphology) are challenging to interpret in a scenario where energetic electrons produced by the population of msPSRs up-scatter ambient radiation fields to gamma-ray energies. On the other hand there is also no support in other wavebands for a scenario where gamma-rays are produced by collisions of target nuclei from the ISM with hadronic cosmic rays originating from a SNR.

Further multi-wavelength observations (X-rays, radio) may help to reveal the nature of HESS~J$1747-248$. In parallel, more sophisticated source models may narrow down the list of applicable emission scenarios.

\section*{Acknowledgements}

The support of the Namibian authorities and of the University of Namibia
in facilitating the construction and operation of H.E.S.S. is gratefully
acknowledged, as is the support by the German Ministry for Education and
Research (BMBF), the Max Planck Society, the French Ministry for Research,
the CNRS-IN2P3 and the Astroparticle Interdisciplinary Programme of the
CNRS, the U.K. Science and Technology Facilities Council (STFC),
the IPNP of the Charles University, the Polish Ministry of Science and 
Higher Education, the South African Department of
Science and Technology and National Research Foundation, and by the
University of Namibia. We appreciate the excellent work of the technical
support staff in Berlin, Durham, Hamburg, Heidelberg, Palaiseau, Paris,
Saclay, and in Namibia in the construction and operation of the
equipment.

\clearpage

\end{document}